\documentclass{iopart}
\expandafter\let\csname equation*\endcsname\relax                  
\expandafter\let\csname endequation*\endcsname\relax 
\usepackage{amsmath}
\usepackage{amssymb}
\usepackage{graphicx, subfigure}
\usepackage{epstopdf}
\usepackage{hyperref}
\usepackage{psfrag}
\usepackage{color}
\usepackage{simplewick}
\usepackage{appendix}
\usepackage{color}
\usepackage{comment}

\allowdisplaybreaks

{\left\lbrace\begin{array}{@{}l@{}}}%
{\end{array}\right.}

\begin{document}

\title{Thermoelectric efficiency of three-terminal quantum thermal machines}

\newcommand{\sns}{NEST, Scuola Normale Superiore, and Istituto Nanoscienze-CNR, I-56126 Pisa, Italy}

\newcommand{\parigi}{Service de Physique de l' \'Etat Condens\'e (CNRS URA 2464), IRAMIS/SPEC, CEA 
	Saclay, 91191 Gif-sur-Yvette, France}

\newcommand{\como}{CNISM and Center for Nonlinear and Complex Systems, Universit\'a degli Studi dell' Insubria,
 via Valleggio 11, 22100 Como, Italy}
	
\newcommand{\milano}{Istituto Nazionale di Fisica Nucleare, Sezione di Milano, via Celoria 16, 20133 Milano, Italy}

\author{Francesco Mazza}
\address{\sns}

\author{Riccardo Bosisio}
\address{\parigi}

\author{Giuliano Benenti}
\address{\como}
\address{\milano}

\author{Vittorio Giovannetti}
\address{\sns}

\author{Rosario Fazio}
\address{\sns}

\author{Fabio Taddei}
\address{\sns}



\begin{abstract}
The efficiency of a thermal engine working in linear response regime in a multi-terminal configuration is discussed. 
For the generic three-terminal case, we provide a general definition of local and non-local transport coefficients: 
electrical and thermal conductances, and thermoelectric powers. 
Within the Onsager formalism, we derive analytical expressions for the 
efficiency at maximum power, which can be written in terms of generalized 
figures of merit. Furthermore, using two examples, we investigate numerically how a third terminal could improve 
the performance of a quantum system, and under which conditions non-local thermoelectric effects can be observed.
\end{abstract}


\maketitle


\section{Introduction}

Thermoelectricity has recently received enormous attention due to the constant demand for new and powerful ways of energy conversion. 
Increasing the efficiency of thermoelectric materials, in the  whole range spanning from macro- to nano-scales, is one of the main challenges, 
of great importance for several different technological applications~\cite{Mahan1997,Majumdar2004,Dresselhaus2007, Snyder2008,Shakouri2011,Dubi2011}. Progress in understanding thermoelectricity at the nanoscale will have important applications for ultra-sensitive all-electric heat and energy transport detectors, energy transduction, heat rectifiers and refrigerators, just to 
mention a few examples. The search for optimisation of nano-scale heat engines and refrigerators has hence stimulated a large body of 
activity, recently reviewed by Benenti {\em et al.}~\cite{Benenti2013}.

While most of the investigations have been carried out in two-terminal setups, thermoelectric transport in multi-terminal devices just begun to be investigated~\cite{Jacquet:2009,Entin-Wohlman:2010,Entin-Wohlman:2012,Jiang:2012,Jiang:2013,Saito:2011,Horvat:2012,Balachandran:2013,Sanchez:2011,Sanchez:2011(b),Sothmann:2012,Sothmann:2012(b),Brandner:2013, Brandner:2013(b),Bedkihal:2013} since these more complex designs  may offer additional advantages. 
An interesting perspective, for instance, is the possibility to exploit a third terminal to ``decouple'' the energy and charge flows 
and improve thermoelectric efficiency~\cite{Entin-Wohlman:2010, Entin-Wohlman:2012, Jiang:2012, Jiang:2013, Sanchez:2011, 
Sanchez:2011(b), Sothmann:2012, Sothmann:2012(b)}. Furthermore, fundamental questions concerning thermodynamic bounds 
on the efficiency of these setups has been investigated\cite{Saito:2011, Horvat:2012, Balachandran:2013, Brandner:2013, Brandner:2013(b)}, 
also accounting for the effects of a magnetic field breaking the time-reversal symmetry\cite{Benenti2011}.  In most of the cases studied so far, however, 
all but two-terminal were considered as mere probes; i.e. no net flow  of energy and charge through them was allowed. 
In other works a purely bosonic reservoir has been used, only exchanging energy (and not charge) current with the 
system~\cite{Entin-Wohlman:2010, Entin-Wohlman:2012, Jiang:2012, Jiang:2013}. 

\begin{figure}
	 \centering
   \includegraphics[width=7cm]{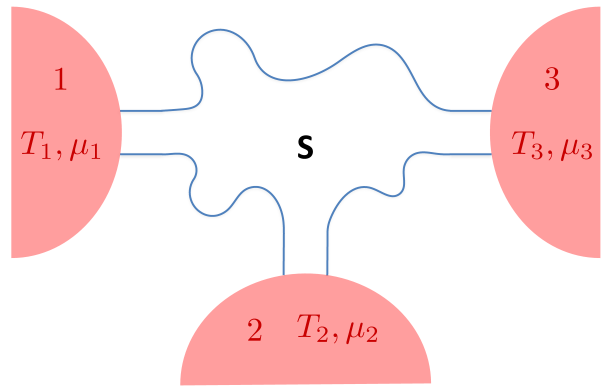}
   \caption{
   Three-terminal thermal machine. A scattering region is connected to 3 different fermionic reservoirs, 
   each of these is able to exchange heat and particles with the system. Reservoir $3$ is taken as the 
   reference for measuring temperature and energy: $T_3 \equiv T$; $\mu_3 = \mu$. The reservoirs $1$ and
   $2$ have small variations in temperature and chemical potential: $(T_i, \mu_i) = (T + \Delta T_i, \mu+\Delta \mu_i)$, 
   $i \in (1,2)$. With ${\bf{S}}$ we denote a generic coherent scattering region.}
   \label{fig:system}
\end{figure}

A genuine multi-terminal device will however offer enhanced flexibility and therefore  it might be useful to improve thermoelectric 
efficiency. A full characterization of these systems is still lacking and motivates us to tackle this problem. Here we focus on the simplest 
instance of three reservoirs, which can exchange both charge and energy current with the system. A sketch of the thermal 
machine is shown in Fig.\ref{fig:system}, where three-terminal are kept at different temperatures and chemical potentials connected through a scattering 
region. Our aim is to provide a  general treatment of the linear response thermoelectric transport for this case, and for this purpose we will discuss
local and non-local transport coefficients. Note that non-local transport coefficients are naturally requested in a multi-terminal setup, since they connect
temperature or voltage biases introduced between two-terminal to 
heat and charge transport among the remaining terminals. 
We will then show that the 
third terminal could be exploited to improve 
thermoelectric performance 
with respect to the two-terminal case. 
We will focus our investigations on the efficiency at maximum 
power~\cite{Yvon:1955,Chambadal:1957,Novikov:1958,CurzonAhlborn:1975,Broeck,Seifert,Esposito2009,Esposito2010,Linke2010,Apertet2012},
i.e. of a heat engine operating under conditions where the output power
is maximized.
Such quantity, central in the field of finite-time 
thermodynamics~\cite{Andresen}, is of great fundamental and practical relevance to understand which systems offer the best trade-off between thermoelectric
power and efficiency. 

The paper is organized as follows. In Section~\ref{sec:3terminals} we 
briefly review the linear response,  Onsager formalism for a generic 
three-terminal setup.  We will discuss the maximum output power and trace a derivation of all the local and non-local 
transport coefficients. In Section~\ref{sec:efficiency} we extend the concept of Carnot bound at the maximum efficiency to the 
three-terminal setup and we derive analytical formulas of the efficiency at maximum power in various cases, depending on the flow 
of the heat currents. These expressions are written in terms of generalized dimensionless figures of merit. 
Note that the expressions derived in Section~\ref{sec:3terminals} and~\ref{sec:efficiency} are based on the properties of the Onsager matrix
and on the positivity of the entropy production. Therefore they hold for non-interacting as well as interacting systems.
This framework will then be applied in Section~\ref{sec:models}  to specific examples of non-interacting systems 
in order to illustrate the salient physical picture.  Namely, we will consider a single 
quantum dot and two dots in series coupled to the three-terminal.  Finally Section~\ref{sec:conclusions} is devoted to the conclusions.


\section{Linear response for 3-terminal systems}
\label{sec:3terminals}

The system depicted in Fig.~\ref{fig:system} 
 is characterized by three energy 
 and three particle currents ($J_{i=1,2,3}^U$ and $J_{i=1,2,3}^N$, 
respectively)   flowing from the corresponding reservoirs, 
which  have to fulfill  the constraints: 
    \begin{eqnarray}\sum_{i=1}^3 J_i^U = 0 \quad (\mbox{Energy conservation})\;, \nonumber \\
 \sum_{i=1}^3 J_i^N = 0 \quad (\mbox{Particle conservation})\;, \label{PARTICLE} 
 \end{eqnarray}  
 (positive values being associated with flows {\it from} the reservoir  {\it to} the system).
In what follows we will assume the reservoir 3 as a reference and
the system to be operating in the linear response regime, i.e. 
set $(T_3,\mu_3) \equiv (T,\mu)$ and write  $(T_j,\mu_j) = (T + \Delta T_j, \mu + \Delta \mu_j)$ 
with $|\Delta  \mu_j|/k_B T\ll 1$ and $|\Delta T_j|/T\ll 1$
for $j=1,2$, and $k_B$ is the Boltzmann constant. 
Under these assumptions  the relation  between currents and biases can then be expressed through the Onsager matrix $L$ of elements $L_{ij}$  via the identity:
\begin{equation}
\begin{pmatrix}
J_1^N \\
J_1^Q \\
J_2^N \\
J_2^Q \\
\end{pmatrix}
=
\begin{pmatrix}
L_{11} & L_{12} & L_{13} & L_{14} \\
L_{21} & L_{22} & L_{23} & L_{24} \\
L_{31} & L_{32} & L_{33} & L_{34} \\
L_{41} & L_{42} & L_{43} & L_{44} \\
\end{pmatrix}
\begin{pmatrix}
X_1^\mu \\
X_1^T \\
X_2^\mu \\
X_2^T \\
\end{pmatrix},
\label{omatrix}
\end{equation}
where $X_{1,2}^\mu = \Delta \mu_{1,2}/T$ and $X_{1,2}^T = \Delta T_{1,2}/T^2$ are the generalized forces, and where 
$J_{1,2}^Q = J^U_{1,2} - \mu_{1,2} J^N_{1,2}$ are the heat currents of the system, the corresponding currents to reservoir 3 being 
 determined from $J_{1,2}^N$ and  $J_{1,2}^Q$ via the conservation laws  of Eq.~(\ref{PARTICLE}). 
In our analysis we take $L$  to be symmetric  (i.e. $L_{ij} = L_{ji}$) by enforcing time reversal symmetry in the problem.
We also  remind that, due to the positivity of the entropy production rate, such matrix has to be  semi-positive definite (i.e. $L\geq 0$) and 
that  it can be used  to describe a two-terminal model connecting (say) reservoir 1 to reservoir 3 by 
 setting $L_{j3}=L_{j4}=L_{3j} = L_{4j} =0$ for all $j$.

\subsection{Transport coefficients}

For a two-terminal model the elements of the  Onsager matrix  $L$ can be related to four quantities 
which gauge the transport properties of the 
system under certain constraints. Specifically these are  the electrical conductance $G$ and the Peltier coefficient $\Pi$ (evaluated
under the assumption that both reservoirs have the same temperature), and  the thermal conductance $K$ and  the thermopower (or Seebeck coefficient) $S$ (evaluated when no net charge current is flowing through the leads).
When generalized to the multi-terminal model these quantities yield to the introduction of non-local coefficients, which describe how transport in a reservoir is influenced by a bias set between two other reservoirs.

\subsubsection{Thermopower}

For a two-terminal configuration the  thermopower relates the voltage $\Delta V$ that develops between the reservoirs to 
their temperature difference $\Delta T$ under the assumption that no net charge current  is flowing in the system, i.e. 
 $S = -\left( \frac{\Delta V}{\Delta T}\right)_{J^N=0}$. 
A  generalization of this quantity to the multi-terminal scenario can be obtained by introducing the  matrix of elements
 \begin{equation}
 \label{eq:thermopower}
 S_{ij} = -\Big(\frac{\Delta \mu_i}{e\Delta T_j} \Big)_{\mbox{\tiny{$\begin{array}{l}
 J^N_k = 0 \; \forall k, \\
  \Delta T_{k} = 0\;  \forall k\neq j 
  \end{array}$}} },
 \end{equation}
with local ($i=j$) and non-local ($i\neq j$) coefficients, $e$ being the electron charge. 
In this definition, which does not require the control of the heat currents, we have imposed that the particle currents in all the leads are zero (the voltages are measured at open circuits) {\em and} that all but one temperature differences are zero (of course this  last condition is not required in a two-terminal model).
It is worth observing  that Eq.~(\ref{eq:thermopower}) differs from other definitions proposed in the literature.
For example in Ref.~\cite{Machon(2013)} a generalization  of the two-terminal thermopower to  a three-terminal system, was proposed 
by setting to zero one voltage instead of the corresponding particle current.
While operationally well defined, this choice does not allow one to easily recover the thermopower of the two-terminal case (in our approach instead this is rather natural, see below). 
Finally in the probe approach presented in 
Refs.~\cite{Jacquet:2009,Saito:2011,Horvat:2012,Balachandran:2013,Brandner:2013,Brandner:2013(b)}  
it was possible to study a multi-terminal device by using an effective two-terminal system only, because the heat and particle currents of the probe terminals are set to vanish by definition. Therefore, within this approach, there are no chances of having non-local transport coefficients.  

In the three-terminal scenario we can use  Eq.~(\ref{omatrix}) to rewrite the elements of the matrix~(\ref{eq:thermopower}).
In particular introducing the quantities
\begin{equation} \label{defL2}
L^{(2)}_{ij;kl} = L_{ik}L_{lj} -  L_{il}L_{kj} ,
\end{equation}
 we get (see ~\ref{app:transport} for details)
\begin{eqnarray}
\label{Sloc}
S_{11} =  \frac{1}{eT}\frac{L^{(2)}_{13;32}}{L^{(2)}_{13;31}}, \qquad 
 S_{22}   = \frac{1}{eT}\frac{L^{(2)}_{14;31}}{L^{(2)}_{13;31}}, \\ 
 S_{12} = \frac{1}{eT}\frac{L^{(2)}_{13;34}}{L^{(2)}_{13;31}} , \qquad \label{Sloc1}
 S_{21} = \frac{1}{eT}\frac{L^{(2)}_{13;21}}{L^{(2)}_{13;31}} ,
\end{eqnarray}
which yields, correctly, $S_{11} = \frac{1}{eT}\frac{L_{12}}{L_{11}}$ as the only non-zero element,
by taking the two-terminal limits detailed at the end of the previous section.


\subsubsection{Electrical conductance}
In a two-terminal configuration the electric conductance describes how the electric current depends upon the bias voltage between the two-terminal under isothermal conditions, i.e. 
$G = \left( \frac{e J^N}{\Delta V} \right)_{\Delta T=0}$. 
The generalization to many-terminal systems is provided by the following matrix: 
 \begin{equation}
 G_{ij} = \Big(\frac{e^2 J_i^N}{\Delta \mu_j} \Big)_{\mbox{\tiny{$\begin{array}{l}
 \Delta T_k = 0 \; \forall k, \\
  \Delta \mu_{k} = 0\;  \forall k\neq j 
  \end{array}$}} }.
 \end{equation}
 Using the three-terminal Onsager matrix~(\ref{omatrix}) we find
 \begin{equation}
 \begin{pmatrix}
 G_{11} & G_{12} \\
 G_{21} & G_{22}
 \end{pmatrix}
 =
\frac{e^2}{T}
 \begin{pmatrix}
 L_{11} & L_{13} \\
 L_{13} & L_{33}
 \end{pmatrix},
 \end{equation}
which, in the two-terminal limit where reservoir 2 is disconnected from the rest, gives $G_{11} = \frac{e^2}{T} L_{11}$ as the only non-zero element.


\subsubsection{Thermal conductance}

The thermal conductance for a two-terminal is the coefficient which 
describes how the heat current depends upon the temperature imbalance $\Delta T$ under the 
assumption that no net charge current is flying through the system, i.e. 
$K = \left( \frac{J^Q}{\Delta T} \right)_{J^N=0}$. In the multi-terminal scenario this generalizes to 
\begin{equation}
K_{ij} = \Big(\frac{J^Q_i}{\Delta T_j} \Big)_{\mbox{\tiny{$\begin{array}{l}
 J^N_k = 0 \; \forall k, \\
  \Delta T_{k} = 0\;  \forall k\neq j 
  \end{array}$}} },
\end{equation}
where one imposes  the same constraints as those used  for the thermopower matrix~(\ref{eq:thermopower}), i.e. no currents
and $\Delta T_k=0$ for all terminals but the $j$-th.
For a  three-terminal case, using Eq.~(\ref{defL2})
this gives 
\begin{equation}
K_{11} = \frac{1}{T^2}\frac{L_{13}L^{(2)}_{12;32}-L_{12}L^{(2)}_{13;32}-L_{11}L^{(2)}_{23;23}}{L^{(2)}_{13;31}},
\end{equation}
\begin{equation}
K_{22} = \frac{1}{T^2}\frac{L_{14}L^{(2)}_{13;43}-L_{13}L^{(2)}_{14;43}-L_{11}L^{(2)}_{34;34}}{L^{(2)}_{13;31}},
\end{equation}
and 
\begin{equation}
K_{12} = K_{21}  = \frac{1}{T^2}\frac{L_{24}L^{(2)}_{13;31}+L_{14}L^{(2)}_{13;23}+L_{34}L^{(2)}_{13;12}}{L^{(2)}_{13;31}}.
\end{equation}
Once more, in the two-terminal limit where the reservoir 2 is disconnected from the rest, the only non-zero element is $K_{11} =  \frac{1}{T^2} \frac{L^{(2)}_{12;12}}{L_{11}}$.


\subsubsection{Peltier coefficient}

In a two-terminal configuration the Peltier coefficient relates the heat current to the charge current under isothermal condition, i.e. 
$\Pi = \left( \frac{J^Q}{eJ^N} \right)_{\Delta T=0}$.
For multi-terminal systems this generalizes to the matrix 
\begin{equation}
\Pi_{ij} = \Big( \frac{J^Q_i}{eJ^N_j} \Big)_{\mbox{\tiny{$\begin{array}{l}
 \Delta T_k = 0 \; \forall k, \\
  \Delta \mu_{k} = 0\;  \forall k\neq j 
  \end{array}$}} },
\end{equation}
which can be shown to be  related to the thermopower matrix~(\ref{eq:thermopower}), 
 through the Onsager reciprocity equations, i.e. $\Pi_{ij}({\bf B} ) =  T S_{ji} (-{\bf B})$ (${\bf B}$  being the magnetic field on the system),\cite{callen,degrootmazur} 
 from which, using Eqs.~(\ref{Sloc}) and (\ref{Sloc1}),  
 one can easily derive for the three-terminal case 
the dependence upon the Onsager matrix  $L$. 
 

\section{Efficiency for 3-terminal systems}
\label{sec:efficiency}

In order to characterize the properties of a multi-terminal system as a heat engine we shall now analyze its efficiency. 
Generalizing the definition for the efficiency of a two-terminal machine~\cite{Benenti2013, callen}, we define 
the steady state heat to work conversion efficiency $\eta$, for a three-terminal machine, as  the power $\dot{W}$  generated by the machine
(which equals to the sum of all the heat currents exchanged between the system and the reservoirs),
divided by the sum of the heat currents {\it absorbed} by the system, i.e. 
\begin{equation} 
\label{eff}
\eta = \frac{\dot W}{\sum_{i_+} J^Q_i}=  \frac{\sum_{i=1}^3 J^Q_i}{\sum_{i_+} J^Q_i}= \frac{ - \sum_{i=1}^2 \Delta \mu_i J_i^N}{\sum_{i_+} J_i^Q} \;,
\end{equation}
where the symbol $\sum_{i_+}$ in the denominator indicates that the sum  is restricted to positive heat currents only, and where in the last expression we used Eq.~(\ref{PARTICLE})
to express $J^Q_3$ in terms of the other two independent currents\cite{note_eta}.

The definition (\ref{eff}) applies only to the case in which $\dot W$ is positive.
Since the signs of the heat currents $J^Q_i$ are not known {\it a priori} (they actually depend on the details of the system), the expression of the efficiency depends on which heat currents are positive.
For the three-terminal system depicted in Fig.~\ref{fig:system} we set for simplicity $T_3<T_2<T_1$ and focus on those situations
where $J_3^Q$ is negative (positive values of $J_3^Q$ being associated with regimes where the machine effectively works as a refrigerator which extract heat from the coldest reservoir of the system). Under these conditions 
the efficiency is equal to
\begin{equation}
\eta_{12}
= \frac{\dot W}{J_1^Q +J_2^Q}\;, \label{CASE2}
\end{equation}
when both $J_1^Q$ and $J_2^Q$ are positive, or 
\begin{equation}
\eta_i
= \frac{\dot W}{J_i^Q}\;, \label{CASE1}
\end{equation}
when for $i=1$ or $2$ only  $J_i^Q$ is positive.

\subsection{Carnot efficiency}\label{carnot}

The Carnot efficiency represents an upper bound for the efficiency and 
is obtained for an infinite-time (Carnot) cycle.
For a two-terminal thermal machine the Carnot efficiency 
is obtained by simply imposing the condition of zero entropy production, namely $\dot S = \sum_i J^Q_i/T_i = 0$.
If the two reservoirs are kept at temperatures $T_1$ and $T_3$ (with $T_3<T_1$), from the definition of the efficiency, Eq.~\eqref{eff}, one gets the two-terminal Carnot efficiency $\eta_C^{II}=1-T_3/T_1$.
The Carnot efficiency for a three-terminal thermal machine is obtained analogously by imposing the condition of zero entropy production, when a reservoir at an intermediate temperature $T_2$ is added.
If $J_1^Q$ only is positive as in Eq.~(\ref{CASE1}), one obtains
\begin{equation}\label{eq:carnot3t_a}
\eta_{C,1}=1-\frac{T_3}{T_1}+\frac{J_2^Q}{J_1^Q}(1-\zeta_{32})=\eta_C^{II}+\frac{J_2^Q}{J_1^Q}(1-\zeta_{32}),
\end{equation}
where $\zeta_{ij} \equiv T_i/T_j$. 
Note that Eq.~(\ref{eq:carnot3t_a})  
 is the sum of the two-terminal Carnot efficiency $\eta_C^{II}$ and a term whose sign is determined by $(1-\zeta_{32})$.
Since $J_1^Q>0$, $J_2^Q<0$ and $\zeta_{32}<1$, it follows that  $\eta_{C,1}$ is always  \emph{reduced} with respect to its two-terminal counterpart $\eta_C^{II}$.
Analogously if only $J_2^Q$ is positive, one obtains
\begin{equation}\label{eq:carnot3t_c}
\eta_{C,2}=\eta_C^{II} - \frac{T_3}{T_1}\left[ \frac{J^Q_1}{J^Q_2} (1-\zeta_{13})-(1-\zeta_{12}) \right] ,
\end{equation}
which again can be shown to be reduced with respect to $\eta_C^{II}$, since $J_1^Q<0$, $J_2^Q>0$, $\zeta_{12}>1$, and $\zeta_{13}>1$.
We notice that this is a hybrid configuration (not a heat engine, neither a refrigerator): the hottest reservoir absorbs heat, while the intermediate-temperature reservoir releases heat.
However, the heat to work conversion efficiency is legitimately defined since generation of power ($\dot W>0$) can occur in this situation.
Finally, if both $J_1^Q$ and $J_2^Q$ are positive as in Eq.~(\ref{CASE2}) one  obtains
\begin{equation}\label{eq:carnot3t_b}
\eta_{C,12}=1-\frac{T_3}{T_1} \left( 1 + \frac{\zeta_{12} -1}{ 1+ \frac{J^Q_1}{J^Q_2} } \right) =
\eta_C^{II} - \frac{T_3}{T_1} \frac{\zeta_{12} -1}{ 1+ \frac{J^Q_1}{J^Q_2} }.
\end{equation}
Since $T_3<T_2<T_1$, the term that multiplies $T_3/T_1$ is positive so that $\eta_{C,12}$ is reduced with respect to the two-terminal case.

It is worth noticing that, in contrast to the two-terminal case, the Carnot efficiency cannot be written in terms of the temperatures only, but it depends on the details of the system.
Moreover, note that 
the Carnot efficiency is unchanged with respect of the two-terminal case
if $T_2=T_3$ in (\ref{eq:carnot3t_a}) or
if $T_2=T_1$ in (\ref{eq:carnot3t_b}).
Indeed, in this situation the quantities $\zeta_{ij}$ are equal to one, making the extra terms in Eqs.~\eqref{eq:carnot3t_a} or \eqref{eq:carnot3t_b} to vanish.

The above results for the Carnot efficiency 
could be generalized to many-terminal systems.
In particular, we conjecture that,
given a system that works between $T_1$ and $T_3$ (with $T_3<T_1$) and adding an arbitrary number of terminals at intermediate temperatures will in general lead to Carnot bounds smaller than $\eta_C^{II}$. On the other hand, adding terminals at higher (or colder) temperatures than $T_1$ and $T_3$ will make $\eta_C$ increase.

Notice that within linear response and via 
Eq.~(\ref{omatrix}) we can express the Carnot efficiencies
(\ref{eq:carnot3t_a})-(\ref{eq:carnot3t_b}) 
in terms of the generalized forces $X_{1,2}^{\mu}$.


\subsection{Efficiency at Maximum Power}
\label{sec:EffMaxPow}
The efficiency at maximum power  is the value of the efficiency evaluated at the values of chemical potentials that maximize the output power $\dot{W}$ of the engine.
In the two-terminal case  the efficiency at maximum power can be expressed 
as~\cite{Broeck}
\begin{eqnarray} 
\eta^{II}(\dot W_{\text{max}}) = \frac{\eta_C^{II}}{2}\frac{ZT}{ZT +2} \;, 
\label{etamax}
\end{eqnarray} 
where $ZT = \frac{G S^2}{K}T$ is  a dimensionless figure of merit which depends upon the transfer coefficient of the system. 
The positivity of the entropy production imposes that such quantity should be non-negative (i.e. $ZT\geq 0$), therefore $\eta^{II}(\dot W_{\text{max}})$
is bounded to reach its maximum value ${\eta_C^{II}}/{2}$ only in the asymptotic limit of 
 $ZT \to \infty$ (Curzon-Ahlborn limit~\cite{Yvon:1955,Chambadal:1957,Novikov:1958,CurzonAhlborn:1975} within linear response~\cite{Broeck}). 

For the three-terminal configuration the output power is a function of the four generalized forces $(X_1^\mu,X_1^T,X_2^\mu,X_2^T)$ introduced in Eq.~(\ref{omatrix}), i.e. 
\begin{equation}
\label{W}
\dot{W} =  - T(J_1^N X_1^\mu +  J_2^N X_2^\mu )\;.
\end{equation}
In the linear regime this  is  a quadratic function  which can be maximized with respect to 
 $X_1^\mu$ and $X_2^\mu$ while keeping $X_1^T$ and $X_2^T$ constant (the existence of a maximum 
 being guaranteed by the the positivity of the entropy production).
 The resulting expression is 
 \begin{equation}
 \dot W_{\text{max}}  = \frac{T^4}{4} (X_1^T, X_2^T) \; M 
\left( \begin{array}{l} 
X_1^T \\  X_2^T
\end{array}  \label{WMAX1}
\right) \;,
\end{equation}
where  $M = \left[\begin{array}{ll}
c & a \\
a & b 
\end{array} \right]$ is a  positive semi-definite matrix, see ~\ref{app:cholesky}, whose elements depends on the Onsager coefficients via the identities 
 \begin{eqnarray}
a &=& G_{12} S_{12}S_{21} + G_{12} S_{11}S_{22} 
	 +G_{22} S_{21}S_{22} \nonumber \\ &&+ G_{11} S_{11}S_{12}\;, \nonumber \\  \nonumber  \\
b &=& G_{11} S_{12}^2 + 2 G_{12}S_{12}S_{22} + G_{22}S_{22}^2\;, \nonumber \\ \nonumber  \\
c &=& G_{11} S_{11}^2 + 2 G_{12}S_{21}S_{11} + G_{22}S_{21}^2\;.
\label{eq:abc}
\end{eqnarray}
Indicating with $\alpha >\beta\geq 0$ the eigenvalues of $M$ we can then further simplify Eq.~(\ref{WMAX1}) by writing it as 
\begin{equation} \label{neweq1}
\dot W_{\text{max}} = (\alpha\cos^2 \theta + \beta \sin^2 \theta) X^2 T^4/4  \;,
\end{equation}
where $X = \sqrt{(X_1^T)^2 + ( X_2^T)^2}$ is the geometric average of system temperatures, while the angle  $\theta$
identify the rotation in the $X_1^T$, $X_2^T$ plane which defines the eigenvectors of $M$. 
We call the parameter 
\begin{equation}
P =  \alpha\cos^2 \theta + \beta \sin^2 \theta
\end{equation}
{\it three-terminal power factor}. It relates the maximum power 
to the temperature difference:
by construction it fulfills the inequality $\beta \leq P \leq \alpha $, the maximum being achieved for $\theta =0$ (i.e. by ensuring that 
$(X_1^T, X_2^T)$ coincides with the eigenvector of $M$ associated with 
its largest eigenvalue $\alpha$). 
Note that in the two-terminal limit we have
$\beta\to 0$, $\alpha\to  G_{11}S_{11}^2 $,
$\cos^2\theta\to 1$, so that the usual 
two-terminal power factor 
$G_{11}S_{11}^2 $ is recovered.

Exploiting Eq.~(\ref{neweq1}) 
we can now write the efficiency at maximum power for the three cases detailed in Eqs.~(\ref{CASE2}) and (\ref{CASE1}).
Specifically we have 
\begin{equation} \label{ETA1}
\eta_1(\dot W_{\text{max}}) = \frac{1}{2 T}  \frac{ \Delta T_1 Z_{11}^c T + \Delta T_2 ( \delta^{-1} Z_{11}^b T + 2 Z_{11}^a T)}{\delta^{-1} (2\tilde y + Z_{11}^a T ) + Z_{11}^c T + 2} ,
\end{equation}
\begin{equation}\label{ETA2}
\eta_2(\dot W_{\text{max}})= \frac{1}{2 T}  \frac{  \Delta T_2 Z_{22}^b T+\Delta T_1 (\delta Z_{22}^c T + 2 Z_{22}^a T) }{\delta (2y + Z_{22}^a T) + Z_{22}^b T + 2},
\end{equation}
and
\begin{equation}\label{ETA12} 
\eta_{12}(\dot W_{\text{max}})= \frac{1}{2 T}  \frac{ \Delta T_1  Z_{12}^c T+  \Delta T_2( 2 Z_{12}^a T + \delta^{-1} Z_{12}^b T) }{\delta^{-1} (2(1+y^{-1}) + Z_{12}^a T + Z_{12}^b T) + 2(1+\tilde{y}^{-1}) +  Z_{12}^a T + Z_{12}^c T } ,
\end{equation}
where we have defined the parameters $\delta = X_1^T/X_2^T = \Delta T_1/\Delta T_2 $, $y = K_{12}/K_{22}$ and 
$\tilde y = K_{12}/K_{11}$ and
 introduced the following generalized $ZT$ coefficients: 
\begin{eqnarray} \label{ZTGEN}
Z^a_{ij} T = \frac{a T}{K_{ij}}, \quad Z^b_{ij} T = \frac{b T}{K_{ij}}, \quad Z^c_{ij} T = \frac{c T}{K_{ij}}\;.
\end{eqnarray} 
The efficiencies (\ref{ETA1}), (\ref{ETA2}) and (\ref{ETA12}) can also be expressed in terms of the corresponding Carnot efficiencies given in  Eqs.~\eqref{eq:carnot3t_a}, \eqref{eq:carnot3t_c} and  \eqref{eq:carnot3t_b}, obtaining the following equations which mimic Eq.~(\ref{etamax}) of the two-terminal case:
\begin{equation} 
\label{eq:eta/etaC1}
\eta_1(\dot W_{\text{max}}) = \frac{\eta_{C,1}}{2} \frac{ Z^b_{11} T + 2 \delta Z^a_{11} T + \delta^2 Z^c_{11} T}{ 2 \tilde y/ y + 4 \delta \tilde y + 2 \delta^2 +  Z^b_{11} T + 2 \delta Z^a_{11} T + \delta^2  Z^c_{11} T} = \frac{\eta_{C,1}}{2} \frac{ \mathcal{Z}_{11}{ T}} { C_1 + \mathcal{Z}_{11}{T}},
\end{equation}
\begin{equation} 
\label{eq:eta/etaC2}
\eta_2(\dot W_{\text{max}}) = \frac{\eta_{C,2}}{2} \frac{  Z^b_{22} T + 2 \delta  Z^a_{22} T + \delta^2  Z^c_{22} T}{ 2 \delta^2 y/\tilde y  + 4 \delta y + 2 +  Z^b_{22} T + 2 \delta  Z^a_{22} T + \delta^2  Z^c_{22} T}= \frac{\eta_{C,2}}{2} \frac{ \mathcal{Z}_{22}{ T}}{ C_2 + \mathcal{Z}_{22} {T}},
\end{equation}
\begin{equation}
\label{eq:eta/etaC12}
\eta_{12}(\dot W_{\text{max}}) = \frac{\eta_{C,12}}{2} \frac{  Z_{12}^b T + 2 \delta Z_{12}^a T + \delta^2 Z_{12}^c T + o(\Delta T_i)}{2  y^{-1} + 4\delta + 2 \delta^2 \tilde y^{-1}  +  Z_{12}^b T + 2 \delta Z_{12}^a T + \delta^2 Z_{12}^c T+ o(\Delta T_i) }\simeq \frac{\eta_{C,12}}{2} \frac{ \mathcal{Z}_{12}{ T}}{ C_{12} + \mathcal{Z}_{12} {T}} ,
\end{equation}
where we have introduced the constants
\begin{eqnarray}
C_1 &=& 2 \tilde y/ y + 4 \delta \tilde y + 2 \delta^2, \label{eq:c1} \\
C_2 &=&  2 \delta^2 y/\tilde y  + 4 \delta y + 2, \label{eq:c2}   \\
C_{12} &=& \tilde y^{-1} + \delta^2 y^{-1} + 2 \delta,  \label{eq:c12}
\end{eqnarray}
and the combinations of figures of merit
\begin{eqnarray} 
\mathcal{Z}_{ij} {T} = ( Z^b_{ij}  + 2 \delta  Z^a_{ij}  + \delta^2  Z^c_{ij} )T\;.
\end{eqnarray} 
Notice also that in writing Eq.~(\ref{eq:eta/etaC12}) we retained only the leading order neglecting contributions of order $\Delta T_i$ or higher. 
 
The above expressions can be used to provide a generalization of the Curzon-Ahlborn limit efficiency for a multi-terminal quantum thermal device. 
Indeed using the Cholesky decompositions on the Onsager matrix, we can prove that the constants $C_1$, $C_2$ defined in  Eqs.~(\ref{eq:c1}), (\ref{eq:c2})  are positive, see ~\ref{CHSEC} for details.
This fact together with the positivity of the quantities $\mathcal{Z}_{ii} {T}$, that we have checked numerically,
implies that the efficiencies $\eta_i(\dot W_{\text{max}})$ are always 
upper bounded by 
half of the associated Carnot efficiencies, i.e.
\begin{eqnarray}\label{CA1}
\eta_i(\dot W_{\text{max}})  \leq  {\eta_{C,i}}/{2} \;,
\end{eqnarray}  
the inequality being saturated when  the generalized $ZT$ coefficients~(\ref{ZTGEN}) diverge.
An analogous conclusion can be reached also for (\ref{eq:eta/etaC12}), yielding
\begin{eqnarray}\label{CA12}
\eta_{12}(\dot W_{\text{max}})  \leq  {\eta_{C,12}}/{2} \;.
\end{eqnarray}  
In this case $C_{12}$ is no longer guaranteed to be  positive due to the presence of  $K_{12}$. Still 
 the inequality~(\ref{CA12}) can be derived
 by observing that the quantities $C_{12}$ and  $\mathcal{Z}_{12} {T}$ entering in the rhs of Eq.~(\ref{eq:eta/etaC12}) have always
the same sign.


\section{Examples}
\label{sec:models}

In this Section we shall apply the theoretical framework developed so far to two specific non-interacting systems attached to three terminals.
Namely, we will discuss the case of a single dot and the case of two coupled dots, in the absence of electron-electron interaction (which cannot be dealt within the Landauer-B\"uttiker formalism).
Our aim is to show that one can easily find situations where the efficiency and output power are enhanced with respect to the two-terminal case.
Furthermore, through the example of the single dot, we find the conditions that guarantee the non-local thermopowers to vanish. 

The coherent flow of particles and heat through a non-interacting conductor can be described by means of the Landauer-B\"uttiker formalism. Under the assumption that all dissipative and phase-breaking processes take place in the reservoirs, the electric and thermal currents are expressed in terms of the scattering properties of the system~\cite{Buttiker:1988,Datta:1995,Imry:1997}. For instance, in a generic multi-terminal configuration the currents flowing into the system from the $i$-th reservoir are:
\begin{align}
&J^N_i=\frac{1}{h}\sum_{j\neq i} \int_{-\infty}^{\infty} dE \, \mathcal{T}_{ij}(E)\, [f_i(E)-f_j(E)], \label{curr1}\\
&J^Q_i=\frac{1}{h}\sum_{j \neq i} \int_{-\infty}^{\infty} dE \, (E-\mu_i) \, \mathcal{T}_{ij}(E)\, [f_i(E)-f_j(E)], \label{curr2} 
\end{align}
where the sum over $j$ is intended over all but the $i$-th reservoir, 
 \emph{h} is the Planck's constant, $\mathcal{T}_{ij}(E)$ is the transmission probability for a particle with energy $E$ to transit from the reservoir $j$ to reservoir $i$, and where finally $f_{i}(E)=\{\exp[(E-\mu_{i})/k_BT_{i}]+1\}^{-1}$ is the Fermi distribution of the particles injected from reservoir $i$ (notice also that we are considering currents of spinless particles).
In what follows we will use the above expressions in the linear response regime where $|\Delta \mu|/k_B T\ll1$ and $|\Delta T|/T\ll 1$, and compute the associated  Onsager coefficients~(\ref{omatrix}), see ~\ref{app:scattering}.


\subsection{Single dot}
\label{sec:singledot}

\begin{figure}[!t]
 \centering
 \includegraphics[keepaspectratio, width=7cm]{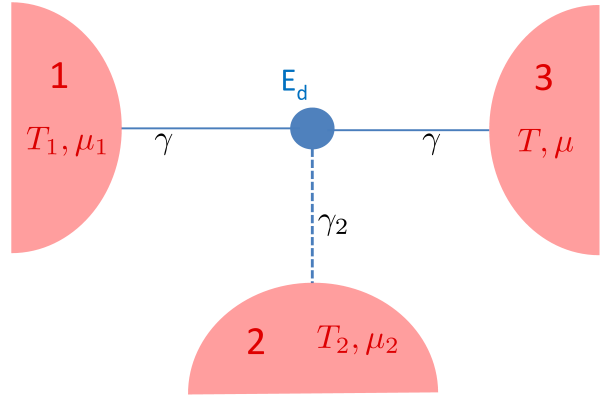}
\caption{(Color online) Sketch of the single dot model used in the numerical simulations: a quantum dot with a single energy level $E_d$ is connected to three fermionic reservoirs $1$, $2$, and $3$. The chemical potential and temperature of the reservoir 3 are assumed as the reference values $\mu$ and $T$. 
The constants $\gamma$ and $\gamma_2$ represent the coupling between the system
and the various reservoirs (see ~\ref{appeSIN} for details). A zero value of $\gamma_2$ corresponds to disconnecting the reservoir 2 from the system: in this regime the
model describes a two-terminal device where reservoirs 1 and 3 are connected through the single dot.}
 \label{fig:Dot}
\end{figure}

In this section we study numerically a simple model consisting of a quantum dot with a single energy level $E_d$, coupled to three fermionic reservoirs, labeled $1$, $2$, and $3$, see Fig.~\ref{fig:Dot}.
For simplicity, the coupling strength to electrodes 1 and 3 are taken equal to $\gamma$, while the coupling strength to electrode 2 is denoted by $\gamma_2$.
In particular we want to investigate how the efficiencies, output powers and transport coefficients evolve when the system is driven from a two-terminal to a three-terminal configuration, that is by varying the ratio $\gamma_2/\gamma$.
The two-terminal configuration corresponds to $\gamma_2=0$ and the third terminal is gradually switched on by increasing $\gamma_2/\gamma$. 
As detailed in \ref{app:scattering},  the transmission amplitudes between each pair of terminals can be 
used to evaluate the Onsager coefficients $L_{ij}$ -- the resulting expression being provided in Eqs.~(\ref{eq:L_ij_2b}).
Once the matrix $L_{ij}$ is known, all the currents flowing through the system, efficiencies, output powers and transport coefficients can be calculated 
within the framework developed in the previous Sections. 
\subsubsection{Efficiencies and maximum power} \label{sec:singeff} 
\begin{figure}[!t]
\centering
 \includegraphics[keepaspectratio, width=10cm]{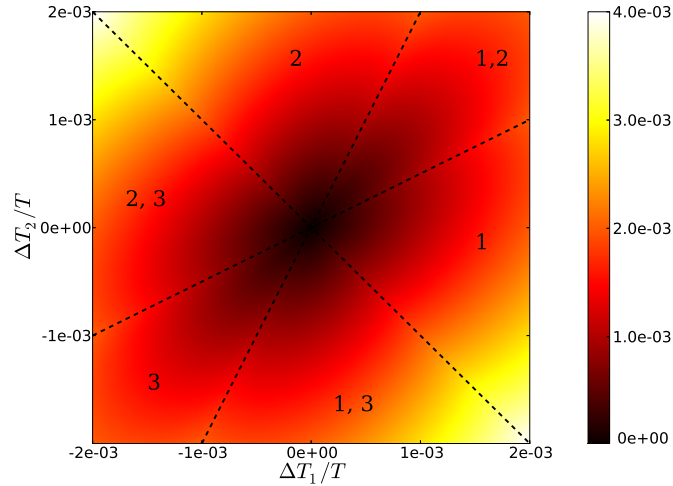}
\caption{(Color online) Carnot efficiency $\eta_C$ (density plot) of the three-terminal system depicted in Fig.~\ref{fig:Dot}, as a function of the gradients of temperature in reservoirs $1$ and $2$ (the chemical potentials $\mu_1$ and $\mu_2$ being chosen to guarantee maximum output power $\dot W$). The coupling with the reservoirs have been set to have  a symmetric configuration with respect to $1$ and $2$ (i.e. $\gamma_2=\gamma$). Note that $\eta_C$ increases linearly along any radial direction defined by a relation $\Delta T_2 = k \Delta T_1$, where $k$ is a constant. In particular, the dashed lines corresponding to $k=0.5$, $k=2$ and $k=-1$ separate different regimes discussed in Sec.~\ref{carnot}.
The numbers in each region identify the reservoirs from which the heat is absorbed. 
Parameter values: $\gamma=0.2 \, k_B T$, $E_d-\mu=2.0 \, k_B T$.}
 \label{fig:Carnot2D}
\end{figure}

In Fig.~\ref{fig:Carnot2D} we show how the Carnot efficiency $\eta_C$ depends on the temperature differences $\Delta T_1$ and $\Delta T_2$, when the chemical potentials are chosen to guarantee maximum output power, i.e., fixing the generalized forces $X^{\mu}_{1,2}$ in order to maximize $\dot{W}$. As we can see, $\eta_C$ increases linearly along any ``radial'' direction defined by a relation $\Delta T_2 = k\, \Delta T_1$, where $k$ is a constant. In particular, the dashed lines corresponding to $k=0.5$, $k=2$, and $k=-1$ separate the different regimes discussed in Sec.~\ref{carnot}: for $-1<k<0.5$ the system absorbs heat only from reservoir $1$ (if $\Delta T_1>0$) or from $2$ and $3$ (if $\Delta T_1<0$); for $0.5<k<2.0$ the system absorbs heat from reservoirs from $1$ and $2$ (if $\Delta T_1>0$) or from $3$ only (if $\Delta T_1<0$); finally, for $k>2$ and $k<-1$ the system absorbs heat only from reservoir $2$ (if $\Delta T_2>0$) or from $1$ and $3$ (if $\Delta T_2<0$). In the case when only one heat flux is absorbed the Carnot efficiency is given by Eq.~(\ref{eq:carnot3t_a}) or Eq.~(\ref{eq:carnot3t_c}), while it is given by Eq.~(\ref{eq:carnot3t_b}) if two heat fluxes are absorbed.

\begin{figure}[!t]
\centering
\includegraphics[keepaspectratio, width=12cm]{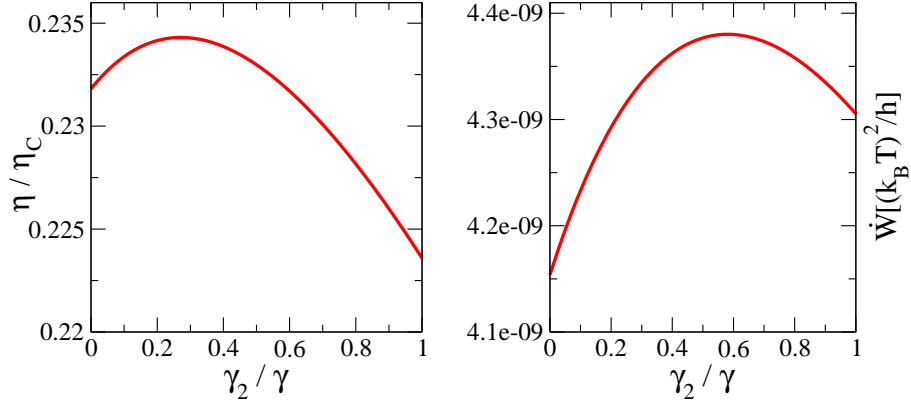}
\caption{(Color online) Left panel: Efficiency $\eta$, normalized over the associated Carnot limit computed as in Sec.~\ref{carnot}, as a function of the coupling to the reservoir $2$. Note that as $\gamma_2/\gamma$ is switched on, the efficiency increases until it reaches a maximum around $\gamma_2 \sim 0.3 \gamma$, and then it decreases. Right panel: Output power $\dot W$ extracted from the system, as a function of the coupling to the reservoir $2$. Parameters: $\gamma=0.1\, k_B T$, $E_d-\mu=2.0 \, k_B T$, $\Delta\mu_1=-\Delta\mu_2=-5\times 10^{-4} \, k_B T$, $\Delta T_1=10^{-3} \, T $, and $\Delta T_2=0$.}
 \label{fig:eta_W_vs_gammaB}
\end{figure}

\begin{figure}[!h]
\centering
\includegraphics[keepaspectratio, width=12cm]{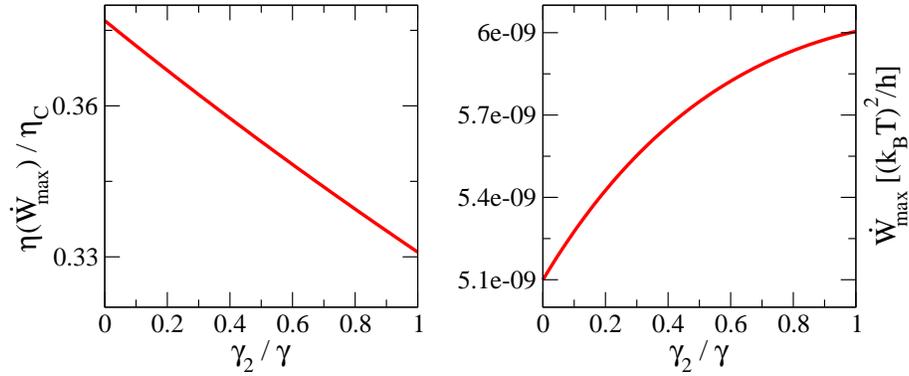}
\caption{(Color online) Left panel: Efficiency at maximum power $\eta (\dot W_{\text{max}})$, normalized over the Carnot limit, as a function of the coupling to the reservoir $2$. 
Right panel: Maximum output power $\dot W_{max}$
extracted from the system, as a function of the coupling to the reservoir $2$. Parameters: $\gamma=0.1\, k_B T$, $E_d-\mu=2.0\, k_B T$, $\Delta T_1=10^{-3 }\, T$, and $\Delta T_2=0$.}
 \label{fig:etaWmax_Wmax_vs_gammaB}
\end{figure}

\begin{figure}[!h]
\centering
\includegraphics[keepaspectratio, width=12cm]{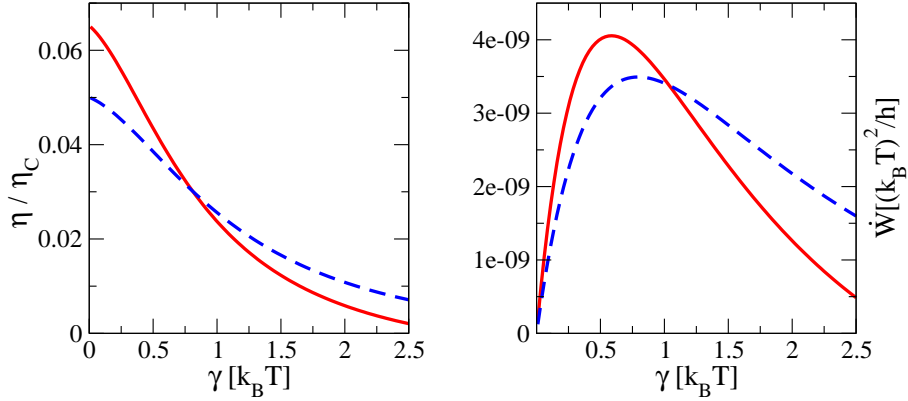}
\caption{(Color online) Left panel: Efficiency $\eta$, normalized over the Carnot limit, as a function of the coupling energy $\gamma$. Right panel: Output power $\dot W$ extracted from the system, as a function of the coupling energy $\gamma$. In both cases, the full red curves correspond to a three-terminal configuration with $\gamma_2=0.5\gamma$, while the dashed blue curve refer to the two-terminal case ($\gamma_2=0$). Parameters: $E_d-\mu=2.0 \, k_B T$, $\Delta\mu_1=-\Delta\mu_2=-10^{-4} \, k_B T$, $\Delta T_1=10^{-3} \, T$, and $\Delta T_2=0$.}
 \label{fig:eta_W_vs_gamma}
\end{figure}

In Figs.~\ref{fig:eta_W_vs_gammaB} and \ref{fig:etaWmax_Wmax_vs_gammaB}, we show how the efficiency Eq.~(\ref{eff}), the output power Eq.~(\ref{W}), the efficiency at maximum output power Eqs.~(\ref{ETA1})-(\ref{ETA12}) and the maximum output power Eq.~(\ref{WMAX1}), vary when the system is driven from a two-terminal to a three-terminal configuration, {\it i. e.} by varying the ratio $\gamma_2/\gamma$.
We set opposite signs for $\Delta\mu_1$ and $\Delta\mu_2$, so that the system absorbs heat only from the hottest reservoir $1$, and $\Delta T_2=0$, in such a way that the Carnot efficiency $\eta_C$ coincides with that of a two-terminal configuration, namely $\eta_C=1-T/T_1$.
Interestingly, we proved that
increasing the coupling $\gamma_2$ to the reservoir $2$ may lead to an improvement of the performance of the system.
In particular, as shown in Fig.~\ref{fig:eta_W_vs_gamma}, the efficiency and the output power can be enhanced \emph{at the same time} 
at small couplings $\gamma_2$, exhibiting a maximum around $\gamma_2\sim 0.3\gamma$ and $\gamma_2\sim 0.6\gamma$, respectively.

In Fig.~\ref{fig:etaWmax_Wmax_vs_gammaB} we show results for the same quantities but at the maximum output power [$\eta(\dot W_{\text{max}})$ and $\dot W_{max}$]. In this case, while $\dot W_{max}$ still increases with $\gamma_2$, the corresponding efficiency decreases approximately linearly.

\begin{figure}[!h]
\centering
\includegraphics[keepaspectratio, width=12cm]{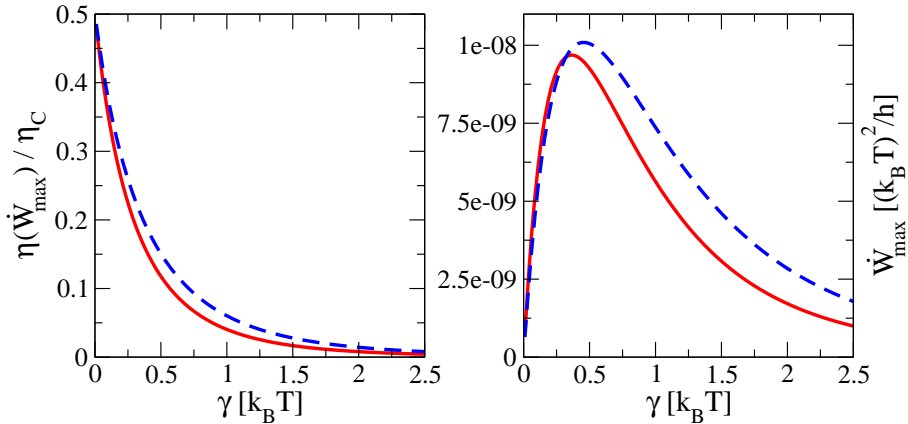}
\caption{(Color online) Left panel: Efficiency at maximum power $\eta (\dot W_{\text{max}})$, normalized over the Carnot limit, as a function of the coupling energy $\gamma$. Right panel: Maximum output power $\dot W_{max}$ extracted from the system, as a function of the coupling energy $\gamma$. In both cases, the full red curves correspond to a three-terminal configuration with $\gamma_2=0.5\gamma$, while the dashed blue curves refer to the two-terminal case ($\gamma_2=0$). Parameters: $E_d-\mu=2.0 \, k_B T$, $\Delta T_1=10^{-3}\, T $, and $\Delta T_2=0$.}
 \label{fig:etaWmax_Wmax_vs_gamma}
\end{figure}

In Figs.~\ref{fig:eta_W_vs_gamma} and \ref{fig:etaWmax_Wmax_vs_gamma} we show the same quantities as in Figs.~\ref{fig:eta_W_vs_gammaB} and \ref{fig:etaWmax_Wmax_vs_gammaB},
but as a function of the coupling $\gamma$ for two values of $\gamma_2$ ($\gamma_2=0$ and $\gamma_2=0.5\gamma$).
From Fig.~\ref{fig:eta_W_vs_gamma} we can see that at small 
$\gamma$ the coupling to a third terminal can enhance both the 
efficiency (for $\gamma\lesssim 0.8k_BT$) and the power (for $\gamma\lesssim k_BT$).
In Fig.~\ref{fig:etaWmax_Wmax_vs_gamma} we note that, both for the
two- and the three-terminal system, the efficiency at
maximum power tends to $\eta/\eta_C=0.5$ in the limit $\gamma\to 0$,
while the output power vanishes. 
For two-terminal the result is well known, since a 
delta-shaped transmission function leads to the divergence of 
the figure of merit $ZT$~\cite{mahansofo,linke1,linke2}.
Correspondingly, the efficiency at maximum 
power saturates the Curzon-Ahlborn bound $\eta/\eta_C=0.5$. 
The same two-terminal energy-filtering argument explains the three-terminal result.
Indeed, we found numerically that for $\gamma\to 0$ 
the chemical potentials optimizing the output power are
such that $\mu_2=\mu_3$. Since also the temperatures are chosen 
so that $T_2=T_3$, we can conclude that terminals $2$ and $3$
can be seen as a single terminal. 


\subsubsection{Thermopowers}

In this section we show analytically that the non-local thermopowers are always zero in this model, while the local ones are equal. 
We consider a general situation, with three different coupling parameters: $\gamma_1 = \gamma$, $\gamma_2 = c\,\gamma$ and $\gamma_3 = d\,\gamma$, with $c\neq d$. Under these assumptions, the transmissions are given at the end of~\ref{app:scattering}. Substituting these expressions in Eqs.~\eqref{Sloc} and \eqref{Sloc1}, we find:
\begin{eqnarray}
S_{11} &=& S_{22} = \frac{1}{eT}\,\frac{L_1}{L_0}, \nonumber \\  
S_{21} & = & S_{12} =0 \;.
\label{eq:1dotthermopower}
\end{eqnarray}
This result is a direct consequence of the factorization of the energy dependence of the transmission probabilities, which are all proportional to the same function 
$\mathcal{T}$, as shown in Eq.~(\ref{eq:T_cd}).
Such factorization allows us to rewrite the Onsager's coefficients 
as in Eq.~(\ref{eq:L_ij_2b}) and derive Eq.~(\ref{eq:1dotthermopower}).  
The fact that the non-local thermopowers, for example $S_{12}$, are zero can be understood as follows. 
Consider first the case in which $T_1=T_2=T_3$ and terminal $2$ behaves as 
a voltage probe. If so, from the condition $J_2^N=L_{31}X_1^\mu
+L_{33}X_2^\mu=0$ we derive $\Delta \mu_2=-(L_{31}/L_{33})\,\Delta\mu_1$. 
Due to the factorization of the energy dependence in the transmissions we
obtain $\Delta\mu_2=(\gamma_1/(\gamma_1+\gamma_3))\Delta \mu_1$. 
Hence, $\Delta \mu_1$ does not depend on the coupling $\gamma_2$. If 
in particular we consider $\gamma_2 = \gamma$, because of the 
symmetry of the system under exchange of the terminal $1$ and $3$
we have $\mu_1 = \mu_3$. 
We can therefore conclude that, independently of the coupling
$\gamma_2$, the probe voltage condition for terminal $2$ 
implies $\Delta\mu_1=0$.
It can be shown that such result remains valid even when 
$\Delta T_1=0$ but $\Delta T_2\ne 0$, as requested in the calculation of 
the thermopower $S_{12}$. As a result, $S_{12}=0$. The same
argument can be repeated for the current $J_1^N$ with the terminal 
$1$ acting as a voltage probe, leading to $S_{21}=0$.

\subsection{Double Dot}
\label{sec:bidot}

\begin{figure}[!t]
 \centering
 \includegraphics[keepaspectratio, width=7cm]{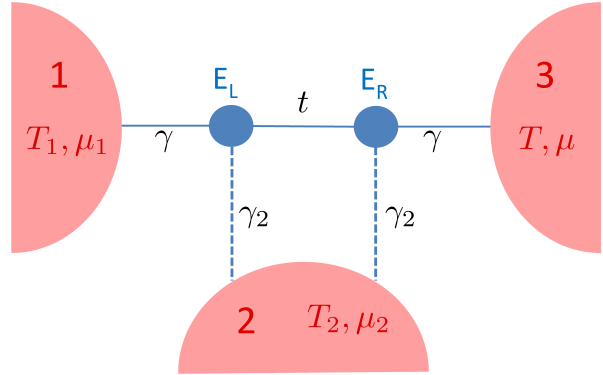}
\caption{(Color online) Sketch of the double dot model used in the numerical simulations: two quantum dots with a single energy level are connected in series to three fermionic reservoirs $1$, $2$ and $3$. The chemical potential and temperature of reservoir $3$ are assumed as the reference values $\mu$ and $T$. A two-terminal configuration is obtained in the case in which the coupling to reservoir $2$ (equal for both the dots) vanishes ($\gamma_2=0$).}
 \label{fig:BiDot}
\end{figure}

Let us now consider a system made of two quantum dots in series, each with a single energy level, coupled to three fermionic reservoirs. This system is described by the Hamiltonian:
\begin{equation}
H = \left[ \begin{array}{cc} E_{L} & -t \\ -t & E_R \end{array} \right].
\end{equation}
We call $t$ the hopping energy between the dots, and we assume that dot $L$ is coupled to the left lead ($1$), dot $R$ is coupled to the right lead ($3$) and that both are coupled to a third lead ($2$) (see Fig.\ref{fig:BiDot}). The self energies describing these couplings are:
\begin{equation}
\Sigma_1 = \left[ \begin{array}{cc} \sigma_{1} & 0 \\ 0 & 0 \end{array} \right],  \quad \Sigma_2 = \left[ \begin{array}{cc} \sigma_{2} & 0 \\ 0 & \sigma_2 \end{array} \right],\quad \Sigma_3 = \left[ \begin{array}{cc} 0 & 0 \\ 0 & \sigma_3 \end{array} \right].
\end{equation}
In the wide-band approximation, we assume that these quantities are energy-independent and they can be written as purely imaginary numbers $\sigma_i=-i\,\gamma_i/2$. The self energies thus become:
\begin{equation}
\Sigma_1 = \left[ \begin{array}{cc} -i\frac{\gamma_1}{2} & 0 \\ 0 & 0 \end{array} \right], \quad \Sigma_2 = \left[ \begin{array}{cc} -i\frac{\gamma_2}{2} & 0 \\ 0 & -i\frac{\gamma_2}{2} \end{array} \right],  \quad  \Sigma_3 = \left[ \begin{array}{cc} 0 & 0 \\ 0 & -i\frac{\gamma_3}{2} \end{array} \right].
\end{equation}
The retarded Green's function of the system is then: 
\begin{flushleft}
\begin{align}
\mathcal{G} = [E\mathbb{I}-H-\Sigma]^{-1} &= \left[ \begin{array}{cc} E-E_{L}-\sigma_1-\sigma_2 & t \\ t & E-E_R-\sigma_3-\sigma_2 \end{array} \right]^{-1}=\cr
&= \frac{1}{\det [\mathcal{G}]}\left[ \begin{array}{cc} E - E_R + i\frac{\gamma_3+\gamma_2}{2} & -t \\ -t & E - E_L + i\frac{\gamma_1+\gamma_2}{2} \end{array} \right],
\end{align}
\end{flushleft}
with
\begin{equation}
\det [\mathcal{G}]=\big(E - E_L + i\frac{\gamma_1+\gamma_2}{2}\big)\big(E - E_R + i\frac{\gamma_3+\gamma_2}{2}\big) - t^2.
\end{equation}
Let us now define the broadening matrices as $\Gamma_i=i(\Sigma_i-\Sigma_i^{\dagger})$:
\begin{equation}
\Gamma_1 = \left[ \begin{array}{cc} \gamma_1 & 0 \\ 0 & 0 \end{array} \right], \quad \quad \Gamma_2 = \left[ \begin{array}{cc} \gamma_2 & 0 \\ 0 & \gamma_2 \end{array} \right], \quad \Gamma_3 = \left[ \begin{array}{cc} 0 & 0 \\ 0 & \gamma_3 \end{array} \right].   
\end{equation}
The matrix of transmission probability $\mathcal{T}_{ij}$ between each pair of reservoirs is then given by the Fisher-Lee formula~\cite{Datta:1995, FisherLee}
\begin{equation}
\mathcal{T}_{ij}=\text{Tr}\,\left[\Gamma_i\,\mathcal{G}\,\Gamma_j\,\mathcal{G}^{\dagger}\right] .
\end{equation}
For the system under consideration we obtain
\begin{equation}
\mathcal{T}_{13}=\frac{\gamma_1\gamma_3}{|\det [\mathcal{G}]|^2}\,t^2,
\end{equation}
\begin{equation}
\mathcal{T}_{12}=\frac{\gamma_1 \gamma_2}{|\det [\mathcal{G}]|^2}\,\left[(E - E_R)^2 + \left(\frac{\gamma_3 + \gamma_2}{2}\right)^2 + t^2\right],
\end{equation}
\begin{equation}
\mathcal{T}_{32}=\frac{\gamma_3 \gamma_2}{|\det [\mathcal{G}]|^2}\,\left[(E - E_L)^2 + \left(\frac{\gamma_1 + \gamma_2}{2}\right)^2 + t^2\right].
\end{equation}
At this point, it is clear that the energy dependence of the transmission matrix cannot be factorized as for the single dot case. This model is hence the simplest in which we can observe finite non-local thermopowers and an increase of both the power and the efficiency of the corresponding thermal machine. We find that the behavior of such quantities as functions of the various parameters is qualitatively very similar to the case of the single dot, thus confirming that a third terminal could improve the performance of a quantum machine.

\begin{figure}[!t]
 \centering
 \includegraphics[keepaspectratio, width=12cm]{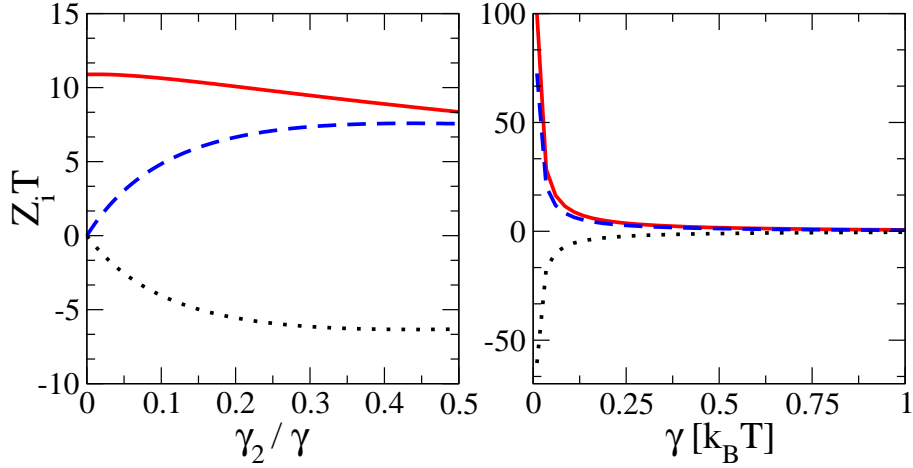}
\caption{Various figures of merit $Z^a_{11}T$ (dotted line), $Z^b_{11}T$ (dashed line) and $Z^c_{11}T$ (full line) as a function of the coupling to the bottom reservoir $\gamma_2$ (left panel), and as a function of the total coupling $\gamma$ (right panel).
Parameter values: $E_L-\mu=-2$ $k_BT$, $E_R-\mu=-20$ $k_BT$,
$\gamma=0.1$ $k_BT$ (left panel) and 
$\gamma_2=0.5$ $k_BT$ (right panel).} 
 \label{fig:ZT_bidot}
\end{figure}

Since in this system all the transport coefficients are different from zero, it is instructive to study the behavior of the generalized figures of merit defined in Eq.~(\ref{ZTGEN}). 
In Fig.~\ref{fig:ZT_bidot} we show, in the configuration with only one positive heat flux ($J^Q_1 >0$), $Z^a_{11}T$ (dotted line), $Z^b_{11}T$ (dashed line) and $Z^c_{11}T$ (full line). We investigate their behavior as a function of the coupling $\gamma_2$ and of the total coupling $\gamma$.
Note that in the two-terminal limit ($\gamma_2\to 0$) $Z^c_{11}T$ reduces to the standard thermoelectric figure of merit $ZT$, while $Z^a_{11}T$ and $Z^b_{11}T$ tend to zero. When we turn on the interaction with the reservoir $2$ (left panel), we notice that the 
figure of merit $Z^c_{11}T$ decreases, while the 
figures of merit $Z^b_{11}T$ and $Z^a_{11}T$ increase their absolute values. From the behavior as a function of the total coupling $\gamma$ we can see that in the limit of $\delta$-shaped transmission function ($\gamma \to 0$), the figures of merit diverge, leading to the Carnot efficiency,
while in the limit of broad transmission window 
($\gamma\to\infty$), all the figures of merit go to zero and we 
recover the case of zero efficiency.
\begin{figure}[!t]
 \centering
 \includegraphics[keepaspectratio, width=12cm]{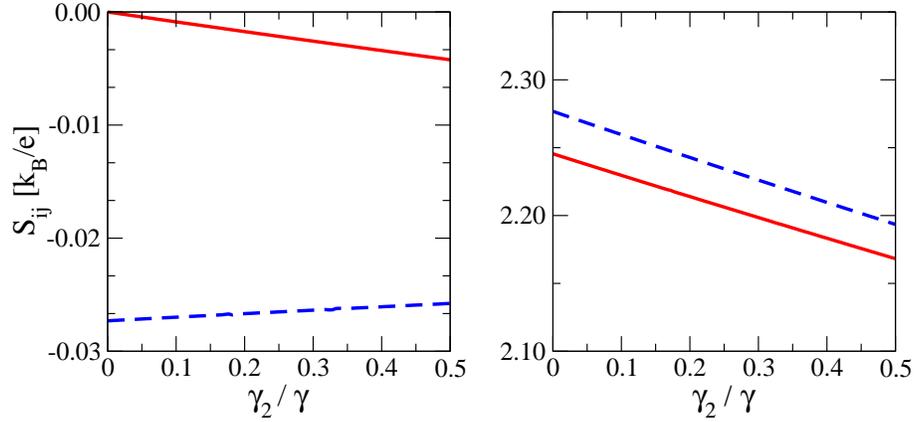}
\caption{(Left panel) Non local thermopowers as a function of the coupling $\gamma_2$ to lead 2. The full red line corresponds to $S_{12} = -\Delta \mu_1/\Delta T_2$, while the dashed (blue) line corresponds to $S_{21} = -\Delta \mu_2/\Delta T_1$. (Right panel) Local thermopowers as a function of the coupling $\gamma_2$ to lead 2. The full (red) line corresponds to $S_{11} = -\Delta \mu_1/\Delta T_1$, while the dashed (blue) line corresponds to $S_{22} = -\Delta \mu_2/\Delta T_2$. Parameter values as in Fig.~\ref{fig:ZT_bidot}.  } 
\label{fig:thermopowers_bidot}
\end{figure}
%


\subsubsection{Thermopowers}

As mentioned before, the fact that the energy-dependence of the transmission matrix for the double dot cannot be factorized is sufficient to guarantee finite non-local thermopowers, as shown in the left panel of Fig.~\ref{fig:thermopowers_bidot}.
As a function of $\gamma_2$, $S_{12}$ starts from zero, while $S_{21}$ starts from a finite value.
This different behavior for the two non-local thermopowers is due to the different role played by $\gamma_2$ in the two cases.
As far as $S_{12}$ is concerned, when we set a temperature difference $\Delta T_2$ in lead 2, a chemical potential difference $\Delta \mu_1$ develops in lead 1 to annihilate the current that flows out of the lead 2. When the coupling $\gamma_2$ goes to zero, that current goes to zero and so does the chemical potential difference $\Delta \mu_1$. This argument does not hold for $S_{21}$, because the temperature difference $\Delta T_1$ is set in lead 1, and $\gamma_2$ does not control the current anymore. Therefore when the coupling $\gamma_2$ approaches zero the current still have a finite value, and so the chemical potential difference $\Delta \mu_2$ needed to annihilate it. 
Furthermore, the local thermopowers are no more equal, as shown in the right panel of Fig.~\ref{fig:thermopowers_bidot}.

\section{Conclusions}
\label{sec:conclusions}

In this paper we have developed a general formalism for linear-response
multi-terminal thermoelectric transport. In particular, we have worked out analytical 
expressions for the efficiency at maximum power in the three-terminal case.
By means of two simple non-interacting models (single- and double-dot), 
we have shown that a third terminal can
be useful to improve the thermoelectric performance of a system with
respect to the two-terminal case. Moreover, we have discussed conditions
under which non-local thermopowers could be observed.
Our analysis could be extended also to cases in which time-reversal 
symmetry is broken by a magnetic field or including
bosonic or superconducting terminals. It is an interesting open 
problem to understand in such instances both 
thermoelectric performance in realistic systems and fundamental bounds 
on efficiency for power generation and cooling.

\section*{Acknowledgements}

The authors would like to acknowledge H. Linke, M. Molteni and S. Valentini for the useful discussions.
This work has been supported by EU project ThermiQ and by MIUR-PRIN: Collective quantum phenomena: from strongly correlated systems to quantum simulators.

\appendix


\section{Calculation of the transport coefficients and thermopowers}
\label{app:transport}

To compute the multi-terminal  thermopowers defined in Eqs.~(\ref{Sloc}) and (\ref{Sloc1}) we have to express one of the temperatures as a function of a thermal current. For example let us start from the inversion between $X_1^T$ and $J_1^Q$. In the Onsager's formalism this can be expressed as:
\begin{equation}
0 =
-\begin{pmatrix}
J_1^N  \\
X_1^T\\
J_2^N \\
J_2^Q \\
\end{pmatrix}
+ \mathcal{L}
\begin{pmatrix}
X_1^\mu\\
J_1^Q\\
X_2^\mu \\
X_2^T \\
\end{pmatrix}
=
\begin{pmatrix}
L &-\mathbb{I}
\end{pmatrix}
A \, A^{-1} 
\begin{pmatrix}
\bf{X}\\
\bf{J}
\end{pmatrix},
\end{equation}
where $A$ is a permutation matrix that switches $X_1^T$ and $J_1^Q$,
$\bf{X}$ and $\bf{J}$ are column vectors with components
$(X_1^\mu,X_1^T,X_2^\mu,X_2^T)$ and 
$(J_1^N,J_1^Q,J_2^N,J_2^Q)$, respectively, and 
$\mathbb{I}$ is the $4\times 4$ identity matrix.
Then we obtain:
\begin{equation}
\begin{split}
0 & = 
\begin{pmatrix}
L &-\mathbb{I}
\end{pmatrix}
A \, A^{-1} 
\begin{pmatrix}
\bf{X}\\
\bf{J}
\end{pmatrix}
= 
\begin{pmatrix}
L &-\mathbb{I}
\end{pmatrix}
A
\begin{pmatrix}
\bf{X^*}\\
\bf{J^*}
\end{pmatrix} 
= \\
& = B {\bf X^*} + C {\bf J^*},
\end{split}
\end{equation}
where $\bf{X^*}$ and $\bf{J^*}$ are the vectors $\bf{X}$ and $\bf{J}$ after the action of $A^{-1}$, that is, with $X_1^T\leftrightarrow J_1^Q$; $B$ and $C$ are the matrices determined by the product 
$
\begin{pmatrix}
L &-\mathbb{I}
\end{pmatrix}
A
$.
We can now define the thermopower from the following equations:
\begin{equation}
{\bf X^*} = -B^{-1} \, C {\bf J^*} \Rightarrow
\begin{pmatrix}
X_1^\mu\\
J_1^Q\\
X_2^\mu\\
X_2^T
\end{pmatrix}
=
\mathcal{L}^{-1} 
\begin{pmatrix}
J_1^N = 0 \\
X_1^T =0 \\
J_2^N =0 \\
J_2^Q
\end{pmatrix}.
\end{equation}
For this choice of the parameters we have inverted, two different thermopowers can be defined, the non local $S_{12}$: 
\begin{equation}
\label{S12}
S_{12} = -\frac{\Delta \mu_1}{e\Delta T_2} = -\frac{1}{eT} \frac{X_1^\mu}{X_2^T} = \frac{1}{eT}\frac{L^{(2)}_{13;34}}{L^{(2)}_{13;31}},
\end{equation}
and the local $S_{22}$:
\begin{equation}
\label{SBB}
S_{22} = -\frac{\Delta \mu_2}{e\Delta T_2} = -\frac{1}{eT} \frac{X_2^\mu}{X_2^T} = \frac{1}{eT}\frac{L^{(2)}_{14;31}}{L^{(2)}_{13;31}}.
\end{equation}
The two-terminal limit in which reservoirs 2 and 3 only are connected 
is obtained after setting in the Onsager matrix 
$L_{ij}=0$ if $i=1,2$ or $j=1,2$.
In this limit, the previous expressions reduce to:
\begin{equation}
\begin{split}
& S_{12}   \to 0, \\
& S_{22}  \to \frac{1}{eT}\frac{L_{34}}{L_{33}}.
\end{split}
\end{equation}
The non-local term goes to zero, while the local one goes to the correct value of the 2-terminal system. The two other terms of these generalized thermopowers are obtained with the inversion of $X_2^T$ and $J_2^Q$. Then we can define $S_{21}$ as the non local quantity, and $S_{11}$ as the local one:
\begin{equation}
\begin{split}
\label{S21-S11}
& S_{21} =-\frac{1}{eT}\frac{X_2^\mu}{X_1^T} = \frac{1}{eT}\frac{L^{(2)}_{13;21}}{L^{(2)}_{13;31}}, \\
& S_{11} =-\frac{1}{eT}\frac{X_1^\mu}{X_1^T} = \frac{1}{eT}\frac{L^{(2)}_{13;32}}{L^{(2)}_{13;31}}.
\end{split}
\end{equation}
In a similar way all the other transport coefficients can be defined, by inverting a generalized force with a current.


\section{Cholesky Decomposition of the three-terminal Onsager matrix}\label{CHSEC}
\label{app:cholesky}

In linear algebra, the Cholesky decomposition~\cite{Cholesky:1998} is a tool which allows to write a Hermitian, positive-definite (or semipositive-definite) matrix $L$ as a product of a lower triangular matrix $D$ and its conjugate transpose $D^{\dag}$:
\begin{equation}
L=D D^{\dag},
\end{equation}
(in particular, if $L$ is \emph{real}, $D^{\dag}$ is simply the transpose of $D$).
It turns out that the sign of some quantities defined throughout this work as combinations of products of Onsager coefficients $L_{ij}$ can be easily studied by using the Cholesky decomposition on the Onsager matrix $L$. As an example, by writing
\begin{equation}
D=
\begin{pmatrix}
\rho_{11} & 0 & 0 & 0 \\
\rho_{12} & \rho_{22} & 0 & 0 \\
\rho_{13} & \rho_{23} & \rho_{33} & 0 \\
\rho_{14} & \rho_{24} & \rho_{34} & \rho_{44} \\
\end{pmatrix},
\end{equation}
it can be shown that the coefficient $b$ and $c$ defined in Eq.~\eqref{eq:abc} are equal to
 \begin{eqnarray}
b &=& \frac{\rho_{14}^2(\rho_{23}^2+\rho_{33}^2)+(\rho_{23}\rho_{24}+\rho_{33}\rho_{34})^2}{T^3 (\rho_{23}^2+\rho_{33}^2)}\;, \nonumber  \\
c &=& \frac{\rho_{22}^2\rho_{23}^2+\rho_{12}^2(\rho_{23}^2+\rho_{33}^2)}{T^3 (\rho_{23}^2+\rho_{33}^2)}\;,
\end{eqnarray}
and therefore are non-negative.
The coefficient 
 \begin{eqnarray}
a &=& \frac{\rho_{12}\rho_{14}(\rho_{23}^2+\rho_{33}^2)+\rho_{22}\rho_{23}(\rho_{23}\rho_{24}+\rho_{33}\rho_{34})}{T^3 (\rho_{23}^2+\rho_{33}^2)} \nonumber 
\end{eqnarray}
instead has undefined sign. 
Still one can prove that it is such that the 
 determinant of the matrix $M$ which appears in Eq.~\eqref{WMAX1} is non-negative. Indeed we have
\begin{equation}
\det({M}) = \frac{(-\rho_{14}\rho_{22}\rho_{23}+\rho_{12}\rho_{23}\rho_{24}+\rho_{12}\rho_{33}\rho_{34})^2}{T^6 (\rho_{23}^2+\rho_{33}^2)}\;,
\end{equation}
which, together with the positivity of $b$ and $c$ entails that $M$ is semi-positive definite.

The same procedure can be used to study the sign of the constants $C_1$, $C_2$ and $C_{12}$ defined in Eqs.~\eqref{eq:c1}, \eqref{eq:c2}, and \eqref{eq:c12}, respectively. As it is shown here below, $C_1$ and $C_2$ are always non-negative, while $C_{12}$ has undefined sign:
 \begin{eqnarray}
C_1 &=& \frac{2[(\delta \rho_{22} \rho_{33}+\rho_{24}\rho_{33}-\rho_{23}\rho_{34})^2+(\rho_{23}^2+\rho_{33}^2)\rho_{44}^2]}{\rho_{22}^2\rho_{33}^2}\;, \nonumber  \\
C_2 &=& \frac{2[(\delta \rho_{22} \rho_{33}+\rho_{24}\rho_{33}-\rho_{23}\rho_{34})^2+(\rho_{23}^2+\rho_{33}^2)\rho_{44}^2]}{(\rho_{24}\rho_{33}-\rho_{23}\rho_{34})^2+(\rho_{23}^2+\rho_{33}^2)\rho_{44}^2}\;, \nonumber  \\
C_{12} &=& \frac{(\rho_{22}\rho_{33}+\delta \rho_{24}\rho_{33}-\delta \rho_{23}\rho_{34})^2+\delta^2(\rho_{23}^2+\rho_{33}^2)\rho_{44}^2}{\rho_{22}\rho_{33}(\rho_{24}\rho_{33}-\rho_{23}\rho_{34})}. \nonumber \\
\end{eqnarray}

\section{Scattering approach in linear response: the Onsager coefficients}
\label{app:scattering}

For a three-terminal configuration, as in the previous sections, we choose the right reservoir $3$ as the reference ($\mu_3 = \mu = 0$, $T_3 = T $), and characterize the problem in terms of the particle/heat currents flowing in linear response between the system and leads $1$ (held at $\mu_1=\mu+\Delta\mu_1$ and $T_1=T+\Delta T_1$) and $2$ (held at $\mu_2=\mu+\Delta\mu_2$ and $T_2=T+\Delta T_2$). 
The Onsager's coefficients are obtained from the linear expansion of 
the currents $J_i^N$ and $J_i^Q$ ($i=1,2$)
given by Eqs.~(\ref{curr1}) and (\ref{curr2}).
They can be written in terms of the transmission probabilities $\mathcal{T}_{ij}$, $i,j\in \{1,2,3\}$ as:
\begin{align}
& L_{11} = \frac{T}{h} \int dE \,\left(-\partial_E f\right) (\mathcal{T}_{12}+\mathcal{T}_{13}),\cr
& L_{12} = \frac{T}{h} \int dE \,\left(-\partial_E f\right) (E-\mu) (\mathcal{T}_{12}+\mathcal{T}_{13})=L_{21},\cr
& L_{13} = \frac{T}{h} \int dE \,\left(-\partial_E f\right) (-\mathcal{T}_{12})=L_{31},\cr
& L_{14} = \frac{T}{h} \int dE \,\left(-\partial_E f\right) (-(E-\mu)\mathcal{T}_{12})=L_{41},\cr
& L_{22} = \frac{T}{h} \int dE \,\left(-\partial_E f\right) (E-\mu)^2 (\mathcal{T}_{12}+\mathcal{T}_{13}),\cr
& L_{23} = \frac{T}{h} \int dE \,\left(-\partial_E f\right) (-(E-\mu)\mathcal{T}_{12})=L_{32},\cr
& L_{24} = \frac{T}{h} \int dE \,\left(-\partial_E f\right) (-(E-\mu)^2\mathcal{T}_{12})=L_{42},\cr
& L_{33} = \frac{T}{h} \int dE \,\left(-\partial_E f\right) (\mathcal{T}_{12}+\mathcal{T}_{23}),\cr
& L_{34} = \frac{T}{h} \int dE \,\left(-\partial_E f\right) (E-\mu) (\mathcal{T}_{12}+\mathcal{T}_{23})=L_{43},\cr
& L_{44} = \frac{T}{h} \int dE \,\left(-\partial_E f\right) (E-\mu)^2 (\mathcal{T}_{12}+\mathcal{T}_{23}),
\label{eq:L_ij}
\end{align}
where $T$ is the temperature, $f$ denotes the Fermi-Dirac distribution at $\mu$, and $\partial_E$ is the partial derivative with respect to the energy.\\

\subsection{Transmission function of a single-level dot}\label{appeSIN}

For a scattering region consisting of a quantum dot with a single energy level, connected to three-terminal, we can express the transmission function as \cite{Buttiker:1988}
\begin{equation}
\mathcal{T}_{ij} = \frac{\Gamma_i\Gamma_j}{(E-E_d)^2+\left(\frac{\Gamma}{2}\right)^2}, \qquad (i\neq j),
\label{eq:T_ij}
\end{equation}
where $\Gamma_i$ is the contribution to the broadening due to the coupling to lead $i$, defined by $\Gamma_i = i (\Sigma_i-\Sigma^{\dagger}_i)$, $\Sigma_i$ being the self-energy of lead $i$. In the wide-band limit approximation, we set $\Sigma_i=-i \, \gamma_i/2$, where $\gamma_i$ does not depend on the energy. Note that this choice leads to the identification $\Gamma_i = \gamma_i$. Furthermore, at the denominator, $\Gamma=\Gamma_1+\Gamma_2+\Gamma_3$ is the total broadening due to the coupling to \emph{all} leads. 
If we denote $\gamma_1=\gamma$, $\gamma_2=c\gamma$, and $\gamma_3=d\gamma$
the couplings to the three leads, we obtain for the transmissions the values
\begin{equation}
\begin{split}
& \mathcal{T}_{12} = \frac{c\gamma^2}{(E-E_d)^2+\frac{(1+c+d)^2}{4}\gamma^2}\equiv c \mathcal{T}, \\
&\mathcal{T}_{13} = \frac{d\,\gamma^2}{(E-E_d)^2+\frac{(1+c+d)^2}{4}\gamma^2}\equiv d \,\mathcal{T}, \\
& \mathcal{T}_{23} = \frac{cd\,\gamma^2}{(E-E_d)^2+\frac{(1+c+d)^2}{4}\gamma^2}\equiv cd \,\mathcal{T}.
\label{eq:T_cd}
\end{split}
\end{equation}
The Onsager coefficients then read as follows:
\begin{align}
& L_{11} = L_0\,(c+d),\cr
& L_{12} = L_1\,(c+d),\cr
& L_{13} = -c\,L_0,\cr
& L_{14} = -c\,L_1,\cr
& L_{22} = L_2\,(c+d),\cr
& L_{23} = -c\,L_1,\cr
& L_{24} = -c\,L_2,\cr
& L_{33} = c \, L_0 \, (1+ d),\cr
& L_{34} = c\, L_1 \, (1+d),\cr
& L_{44} = c\, L_2  \, (1+d),
\label{eq:L_ij_2b}
\end{align}
with $ L_{n} = \frac{T}{h} \int dE \,\left(-\partial_E f\right) (E-\mu)^n \mathcal{T}$.

The numerical data shown in Sec.~\ref{sec:singledot} are obtained
for $d=1$, i.e. for $\gamma_1=\gamma_3=\gamma$. The two-terminal 
configuration corresponds to $c=0$ ($\gamma_2=0$), while the coupling
to terminal 2 is switched on progressively by increasing $c$.

\newpage
\section*{References}
\bibliographystyle{iopart-num}

\begin{thebibliography}{88}
\bibitem{note_eta}
Note that Eq.~(\ref{eff}) can be easily generalized to $M$ terminals,
after appropriate change of the numerator.
By setting for instance $(T_M,\mu_M)=(T,\mu)$,
$\Delta T_i=T_i-T$, and $\Delta\mu_i=\mu_i-\mu$ 
($i=1,...,M-1$), the output power reads as follows:
\begin{equation}
\dot{W}=\sum_{i=1}^M J_i^Q=-\sum_{i=1}^{M-1}
\Delta \mu_i J_i^N.
\end{equation}
\end{thebibliography}


\end{document}